\documentclass[amsmath,amssymb,aps,superscriptaddress,floatfix,twocolumn,pra]{revtex4-1}
\usepackage{amsmath}
\usepackage{amsfonts}

\usepackage{graphicx}
\usepackage[dvipsnames]{xcolor}
\usepackage{microtype}

\usepackage{multirow,bigdelim,booktabs,enumitem}
\usepackage{subfigure, placeins}

\usepackage{physics}

\usepackage[urlcolor=blue, colorlinks=true, bookmarks=false]{hyperref}

\newcommand{\Da}{D_1}
\newcommand{\Db}{D_2}
\newcommand{\Dc}{D_3}
\newcommand{\da}{d_1}
\newcommand{\db}{d_2}
\newcommand{\dc}{d_3}
\newcommand{\At}{A_t}
\newcommand{\Ab}{A_b}
\newcommand{\at}{a_t}
\newcommand{\ab}{a_b}
\newcommand{\ns}{\mathrm{ns}}
\newcommand{\epsbar}{\bar{\epsilon}}
\newcommand{\Tone}{T_{1}}
\newcommand{\hatX}{\hat{X}}
\newcommand{\us}{\mu\mathrm{s}}

\begin{document}

\title{Real-time decoding of stabilizer measurements in a bit-flip code}

\author{Diego~Rist\`e}
\email{diego.riste@raytheon.com}
\author{Luke~C.~G.~Govia}
\author{Brian~Donovan}
\altaffiliation{Current affiliation: Systems \& Technology Research, Woburn, MA, USA}
\author{Spencer~D.~Fallek}
\author{William~D.~Kalfus}
\affiliation{Raytheon BBN Technologies, Cambridge, MA, USA}
\author{Markus~Brink}
\author{Nicholas~T.~Bronn}
\affiliation{IBM T. J. Watson Research Center, Yorktown Heights, NY, USA}
\author{Thomas~A.~Ohki}
\affiliation{Raytheon BBN Technologies, Cambridge, MA, USA}

\date{\today}

\begin{abstract}
  Although qubit coherence times and gate fidelities are continuously improving, logical encoding is essential to achieve fault tolerance in quantum computing.
  In most encoding schemes, correcting or tracking errors throughout the computation is necessary to implement a universal gate set without adding significant delays in the processor.
  Here we realize a classical control architecture for the fast extraction of errors based on multiple cycles of stabilizer measurements and subsequent correction.
  We demonstrate its application on a minimal bit-flip code with five transmon qubits, showing that real-time decoding and correction based on multiple stabilizers is superior in both speed and fidelity to repeated correction based on individual cycles. Furthermore, the encoded qubit can be rapidly measured, thus enabling conditional operations that rely on feed-forward, such as logical gates.
  This co-processing of classical and quantum information will be crucial in running a logical circuit at its full speed to outpace error accumulation.
\end{abstract}

\maketitle

\section{Introduction}

Fault-tolerant quantum computation offers the potential for vast computational advantages over classical computing for a variety of problems~\cite{Nielsen10}.
The implementation of quantum error correction (QEC) is the first step towards practical realization of any of these applications.
This typically requires detecting the occurrence of an error by performing a stabilizer measurement, followed by either a corrective action on the physical device (active QEC), or a frame-update in software (passive QEC) \cite{Knill05,DiVincenzo07,Chamberland18}.
In either case, these checkpoints must occur regularly to protect a quantum state throughout the computation. To do that, one needs to rapidly measure the stabilizers with high fidelity and without disrupting the encoded qubit. In spite of these challenges, recent progress has been made in repetitive stabilizer measurements across a diverse range of physical architectures including trapped ions~\cite{Schindler11, Negnevitsky18}, superconducting qubits~\cite{Kelly15, Ofek16, Hu19, Andersen19, Bultink19}, and defects in diamond~\cite{Cramer16}.

Active feed-forward control is useful not only for active error correction~\cite{Paetznick13,Jochym-OConnor14,Anderson14,Yoder16}, but also for other QEC schemes employing state injection and magic-state distillation~\cite{DiVincenzo07,Fowler12}, both of which may be used in implementations of a universal logical gate set. In all cases, determining the appropriate control and implementing it in \emph{real time} with minimal latency is a particularly attractive capability for efficient error correction techniques~\cite{Chamberland18}.

Furthermore, when the stabilizer measurements cannot be trusted as they themselves are error-prone, one can introduce a decoder that uses information from multiple rounds of stabilizer measurements~\cite{Tomita14} or from the spatial connectivity of the device~\cite{Heim16} to determine the appropriate correction. Unfortunately, performing the decoding calculation in software at a high level of the hardware stack hinders low-latency correction and fast feed-forward control due to the communication and computation overhead~\cite{Kelly15}.

In this Letter, we overcome this bottleneck by performing both QEC decoding and control with custom low-latency hardware, which acts as a classical co-processor to our quantum processor. We demonstrate repeated active correction as well as real-time decoding of multi-round stabilizer measurements. We show that the decoding strategy successfully mitigates stabilizer errors and identifies the encoded state with a latency far below the qubit coherence times, while matching the results obtained by post-processing on a conventional computer.

\section{Setup}
\label{sec:setup}
\begin{figure*}
  \includegraphics[width=1.8\columnwidth]{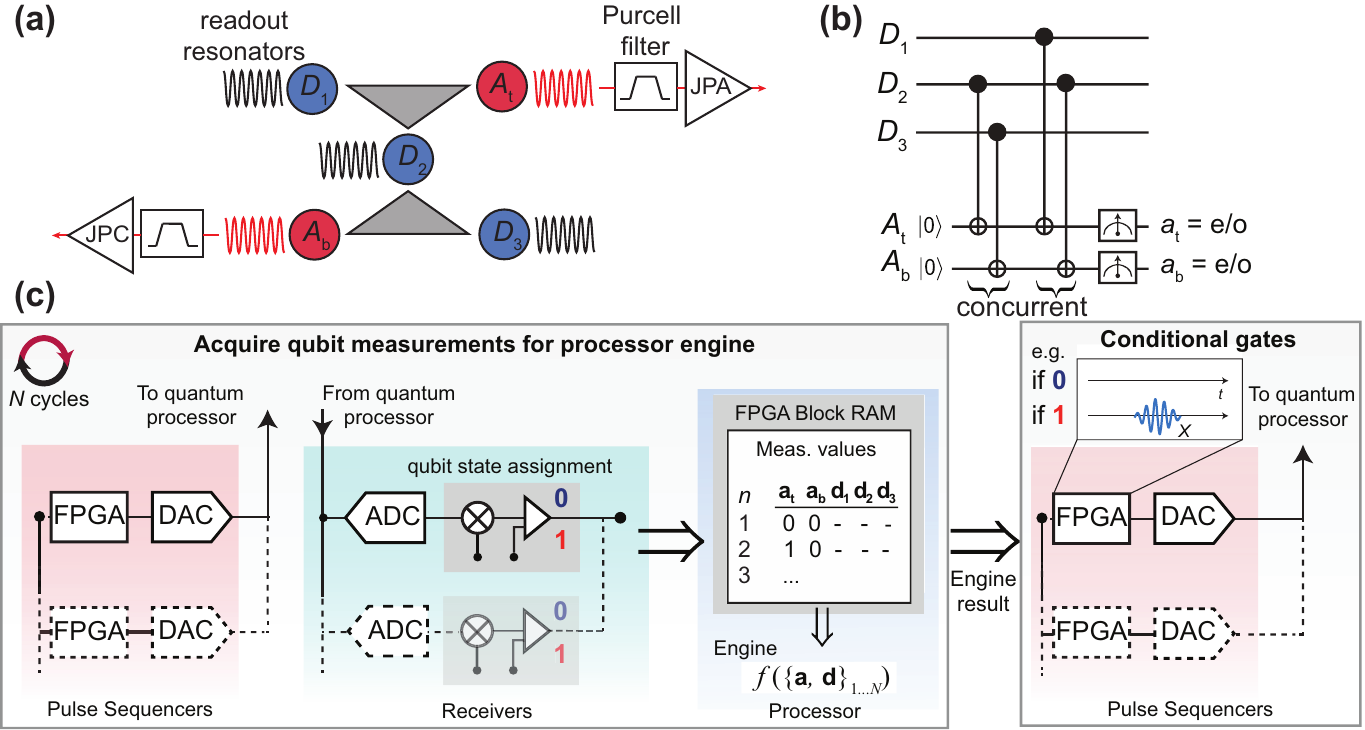}
  \caption{Stabilizer measurements on a five-transmon device. (a) Schematics of the device implementing the bit-flip code with the data qubits $\Da, \Db, \Dc$ and the ancilla qubits $\At, \Ab$.
  Triangles represent the bus resonators coupling the qubits at their vertices. A Josephson Parametric Amplifier~\cite{Hatridge11} (Converter~\cite{Abdo11}) enhances the readout of $\At$ ($\Ab$).
  (b) Gate and measurement sequence for one round of stabilizer measurements. CNOT gates map the parity of $\Da$ - $\Db$ ($\Db$ - $\Dc$) onto $\At$ ($\Ab$) and are applied concurrently two at a time.
  (c) Simplified setup diagram highlighting the closed loop central to active error correction and decoding. For QEC cycles $n \leq N$, the Processor stores the stabilizer results $\{\at,\ab\}_n$ acquired by the Receivers. When $n=N$, it executes a custom function (here Decoder) and broadcasts the result back to the Pulse Sequencers for conditional gates (here $\hat{X}$).
  The same framework is used to execute a logical data measurement, where the Majority function is applied on a single acquisition $\{\da,\db,\dc\}$.  Further detail is provided in Sec.~\ref{sec:setup}.}
  \label{fig:1}
\end{figure*}

For our demonstration, we implement a three-qubit code that corrects bit-flip errors ($\hat{X}$), and is sufficient to encode one logical bit of classical memory. We use an IBM five-transmon device similar to \emph{ibmqx2}~\cite{Qiskit19, Riste17} (Fig.~\ref{fig:1}), of which three transmons ($\Da$, $\Db$, $\Dc$) are used as data qubits, and two ($\At$, $\Ab$) as ancilla qubits to measure the stabilizers. Each qubit is coupled to a dedicated resonator for readout and control.
Additional resonators dispersively couple $\Da$, $\Db$ with $\At$ and $\Db$, $\Dc$ with $\Ab$.

We perform CNOT gates between data and ancilla qubits by using a sequence of single-qubit gates and a $ZX_{90}$ rotation driven by the cross-resonance interaction~\cite{Chow14}. By applying two CNOT gates in succession controlled by two different data qubits with a single ancilla as the target, the parity of the data qubit pair is mapped onto the ancilla state.
The same protocol is applied simultaneously to both data qubit pairs, with the shared qubit $\Db$ interacting first with $\At$, then with $\Ab$ (Fig.~\ref{fig:1}b). The ancilla measurement result $a_{t,b} = 0 (1)$ ideally corresponds to even (odd) parity for the corresponding pair.
We refer to the complete sequence comprised of $4$ CNOT gates and ancilla measurement as a single error correction cycle. The result of each cycle (the measurements ${\at, \ab}$) is a syndrome which identifies which data qubit (if any) has most likely been subjected to an $\hat{X}$ error.

Key to preserving a logical state is the capability of repeating such stabilizer measurements~\cite{Kelly15, Andersen19, Bultink19, Negnevitsky18},
which has two technical requirements.
First, the two ancilla qubits must be reused at every cycle, either by resetting them to the ground state~\cite{Riste12b, Egger18} or by tracking their state. For either reset or state-tracking by measurement, we need to ensure that the readout process is nondestructive, i.e., the result is consistent with the qubit state at the end of the measurement. This sets an upper limit to the allowable photon number, and therefore to the readout fidelity~\cite{SOM}.
Second, the readout cavities must be depleted of photons before starting the new cycle to prevent gate errors. To accelerate the cavity relaxation to its steady-state (near vacuum), we employ the CLEAR technique~\cite{McClure16} for the slower resonator coupled to $\Ab$ (Table~S2), reducing its average photon population to $<0.1$ in $600~\ns$.
Altogether, we measure a single-round joint stabilizer readout fidelity of $0.61$, averaged over the 8 computational states of the data qubits (see Supplemental Material Fig.~S2).

An integral part of our experiment is the interdependence of qubit readout and control, mediated by fast processing of the measurement results. For this purpose, we use a combination of custom-made and off-the-shelf hardware consisting of Pulse Sequencers, Receivers, and a Processor (see Fig.~\ref{fig:1}c), all based on field-programmable gate arrays (FPGAs).
In particular, the storage capability of the Processor FPGA enables expansion beyond \emph{one-time} feedback protocols~\cite{Ofek16, Ryan17, Salathe18}, where conditional actions rely on single, or joint but simultaneous measurements.
During each QEC cycle, digital-to-analog converters (DACs) in the Pulse Sequencers produce a pre-programmed series of gate and measurement pulse envelopes.
Each returning readout signal is captured by a Receiver channel via an analog-to-digital converter (ADC), where it is integrated and compared against a calibrated threshold to determine the qubit state~\cite{SOM, Ryan17}.
The Processor collects all the digitized results and stores them in memory.
After a preset number $N$ of cycles have been executed, the Processor
feeds the stored values to an internal custom calculation engine. The engine function result is broadcast to the Pulse Sequencers, which conditionally apply a corresponding set of gates. The overall latency to store and process the classical data and to issue a conditional pulse is $590~\ns$, a small fraction of coherence times~\cite{SOM}.

We will explore three distinct approaches to the bit-flip code. In all cases, we first prepare the logical excited state $|111\rangle$. Next, we apply one of the following schemes: i) \emph{uncorrected}, in which the cycle is performed but no correction is applied to the data qubits based on the syndrome measurements, and the ancilla qubits are not reset; ii) \emph{repeated error correction} (REC), in which a correction gate is conditionally applied to the data qubits after each error correction cycle based on the syndrome result, and the ancillas are reset; iii) \emph{decoder error correction} (DEC), in which $N$ cycles are performed without ancilla reset or corrective gates and the set of syndromes from all $N$ cycles are used to determine and apply the optimal correction via a decoder~\cite{Tomita14}.
To assess how well each code has protected the prepared state after a desired number of cycles, we perform a logical data measurement. This involves measuring the constituent physical data qubits and computing the majority function over the digitized results $\{d_1, d_2, d_3\}$. In cases i) and ii), the majority function is calculated offline. In case iii), the Processor computes both the decoding and majority functions sequentially~\cite{SOM}, making the result available for further conditional operations.

\section{Results}

\begin{figure}
  \includegraphics[width=\columnwidth]{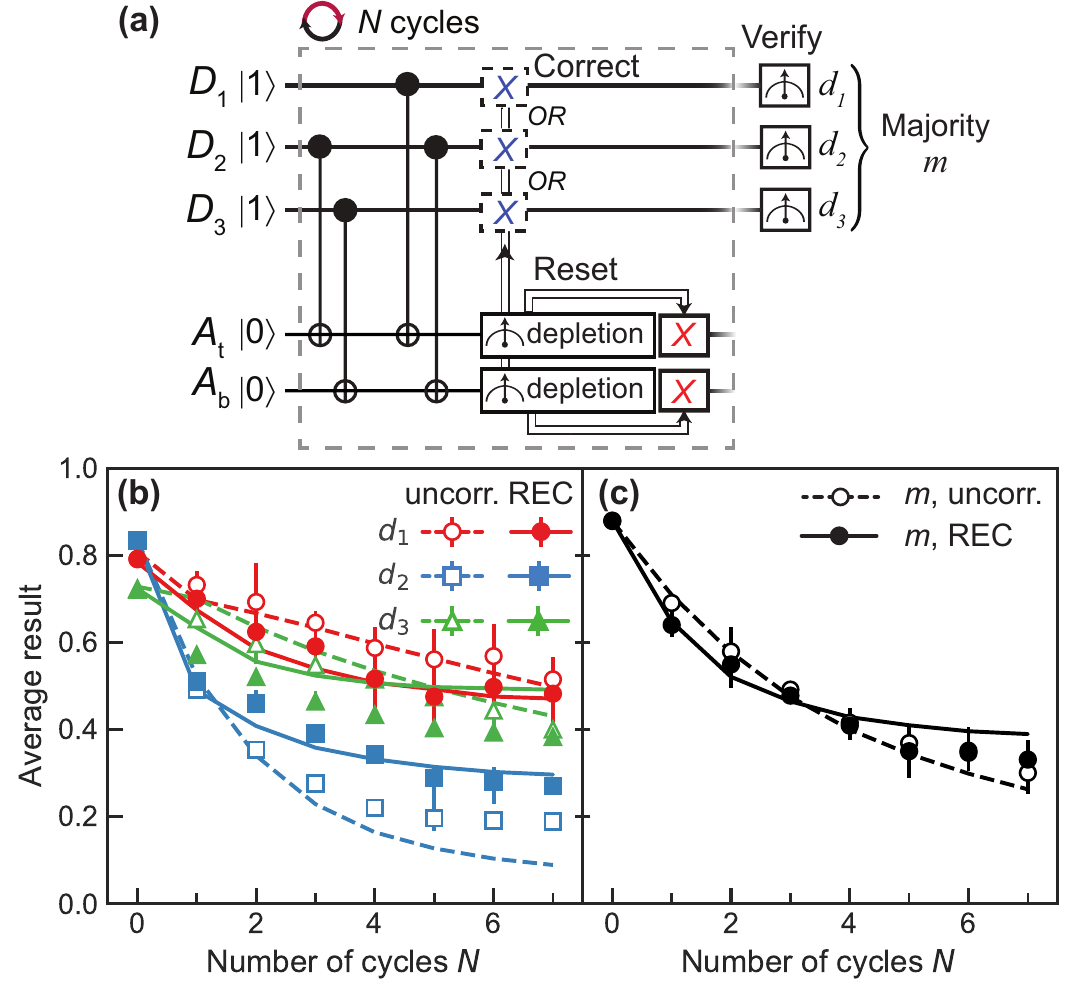}
  \caption{Real-time repeated error correction with ancilla reset (REC).  (a) Gate sequence for independent cycles of active error correction. (b) Average digitized result for each of $D_1$ (red), $D_2$ (blue), and $D_3$ (green) measurements after initialization in $\ket{111}$ and $N$ cycles of (a) (full symbols).
  The results are compared to those obtained with the same circuit, but where data correct and ancilla reset operations are omitted (empty symbols). (c) Average majority vote $m$ for the data in (b). The closed-loop circuit (full symbols) does not improve over open-loop (empty). In both (b-c), solid (dashed) curves are obtained from the model for the corrected (uncorrected) case~\cite{SOM}. Error bars correspond to the range over 5 repetitions of the experiment, consisting of 3000 shots each.}
  \label{fig:2}
\end{figure}

We begin by comparing the REC protocol to the uncorrected case (Fig.~\ref{fig:2}).
For REC, there is a one-to-one relation between the two-bit value syndrome $\{\at, \ab\}$ and one of the 3 possible corrective $\hatX$ gates (in blue in Fig.~\ref{fig:2}a), or no gate at all.
The same syndrome value is used to actively reset the ancilla qubits for use in subsequent cycles (Fig.~\ref{fig:2}b).

When the correction is based on a single round of stabilizer measurements, false positives (largely due to CNOT gate and ancilla readout errors) immediately propagate to the data qubits. These errors dominate in the case of $\da$ and $\dc$, whose average values decay faster with the number of cycles than without active correction (Fig.~\ref{fig:2}b).
Conversely, the larger intrinsic error per cycle for $\db$ (due to its shorter $\Tone$) is partially compensated by the protocol.
Overall, this gain nearly balances out the errors introduced by the active error correction, as shown by comparing the results of the majority function  (Fig.~\ref{fig:2}b).
REC can be thought of as repeated one-time feedback, where the Processor storage and calculation engine are unused. The added latency is considerable:  for each cycle, the stabilizer results are aggregated by the Processor and forwarded to the Pulse Sequencers ($400~\ns$), followed by correction and reset operations ($160~\ns$).

\begin{figure}
 \includegraphics[width=\columnwidth]{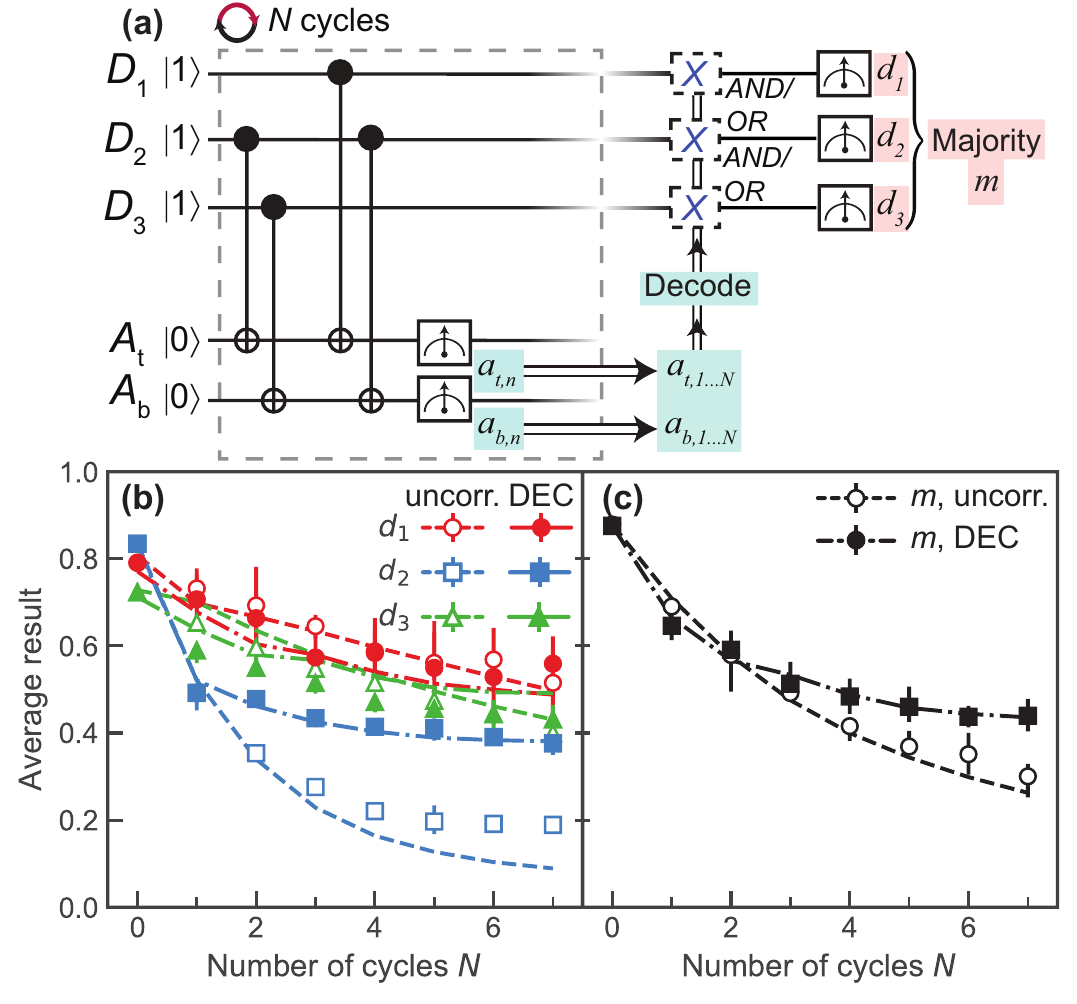}
 \caption{Real-time decoder-based error correction (DEC) and logical measurement. (a) Gate sequence with active correction, based on $N$ rounds of stabilized measurements, followed by simultaneous data qubit measurements and majority function. The sequence in each cycle is identical to the uncorrected case in Fig.~2 (including the cavity depletion, not shown). The shaded areas highlight the real-time processing.
 (b) Average results $\da$, $\db$, $\dc$ following the conditional $\hatX$ gates in (a) (solid symbols), compared to the uncorrected case (same as Fig.~\ref{fig:2}b) (empty). Model curves for real-time correction are dash-dotted. (c) Average majority vote $m$ for the data in (b).}
 \label{fig:3}
\end{figure}

Improvements in logical state protection are achieved by correlating multiple stabilizer measurements using the DEC protocol. In Ref.~\cite{Kelly15}, a simplified minimum-weight perfect matching decoder~\cite{Tomita14} was used to post-process the syndrome results and differentiate between true data bit flips and false positives. We apply the same method (Fig.~\ref{fig:3}), but with the crucial difference that the results are processed in real time.
Specifically, the Processor acquires stabilizer measurement results for $N$ cycles and uses the engine to decode them into the appropriate set of $\hatX$ gates using a precomputed lookup table. These corrections are then applied by the Pulse Sequencers on the data qubits. Finally, the data qubits are measured as in Fig.~\ref{fig:2}, with the majority function also computed on the Processor. Whereas for $N\leq 2$ the decoder is ineffective -- as there are not enough records to identify ancilla readout errors -- a gap emerges at larger $N$ (Fig.~\ref{fig:3}c) in favor of the decoder.
Furthermore, this scheme eliminates the per-cycle latency cost; the latest ancilla results can be processed while the next cycle is executing. The total additional latency becomes fixed at $1300~\ns$ ($590~\ns$ for each Processor engine call, $120~\ns$ for corrective gates), approximately equal to that accrued over 2 REC cycles (Table ~\ref{table:0}).

\begin{figure}
\centering
 \includegraphics[width=0.75\columnwidth]{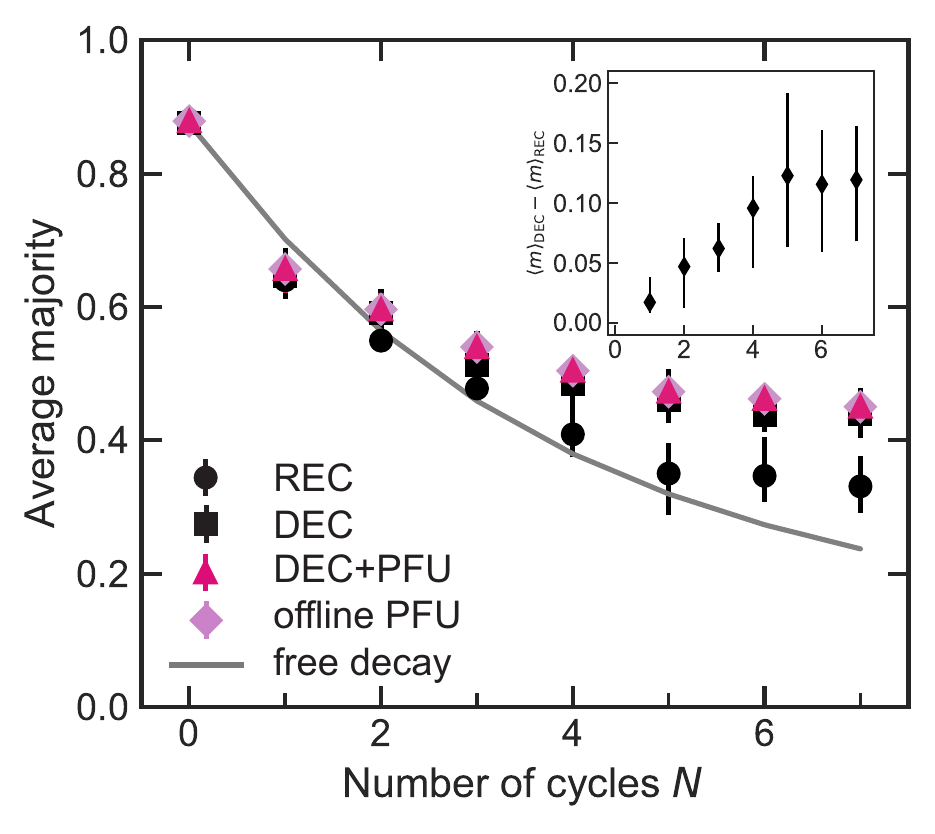}
 \caption{Comparison between different QEC schemes.
 Measured average majority result $m$ for REC (circles, from Fig.~\ref{fig:2}) and DEC with final active correction (squares, Fig.~\ref{fig:3}).
 The results obtained by replacing the correction in DEC with a PFU (triangles) match those obtained by post-processing (diamonds) the uncorrected data using the same decoder.
 In all DEC cases and for $N>1$, $m$ stays above the result expected from an equivalent idle time (gray curve).
 Inset: difference between the average majority result for DEC with PFU and REC.}
 \label{fig:4}
\end{figure}

\begin{table}
\begin{ruledtabular}
\begin{tabular*}{\columnwidth} {@{}ccc@{}}
Protocol & Latency/cycle (ns) & Fixed latency (ns)\\
\toprule

REC & 560 & $>1\mathrm{e}6$ \\

DEC & 0 & 1300  \\

DEC + PFU & 0 & 1180  \\
\end{tabular*}
\end{ruledtabular}
\caption{\label{table:0} Additional latency compared to the uncorrected protocol. The latency includes the time for classical processing, as well as active correction, where applicable.}

\end{table}

In an effort to minimize errors by avoiding unnecessary quantum gates, it is important to move as many operations as possible to the classical hardware.
In this case, we note that measuring the data qubits immediately after conditional $\hatX$ gates is equivalent to inverting the classical measurement result. Therefore, in a final experiment (Fig.~\ref{fig:4}, triangles) we dispense with the active correction and instead filter the $d_i$ results based on the decoder output. This corresponds to a Pauli-frame update (PFU)~\cite{Knill05} applied just before the measurement.
The slight reduction in latency ($120~\ns$) and error rate due to the absence of these pulses consistently achieves $1-2\%$ improvement for all $N$ over the actively corrected case. The results match those obtained by post-processing all the data in software (diamonds), confirming that the fast classical loop works as expected.

Finally, we evaluate the decoder against the majority result we would obtain by replacing the QEC gates and measurements with an idle time of equal duration. The result shows that, for all $N>1$, DEC has a higher success probability of determining the initial state compared to free decay (Fig.~\ref{fig:4}, gray curve).

\section{Conclusions}
Although the experiment ends with the measurement of the data qubits, the readily available majority result may be used to condition additional operations on a second encoded qubit, such as for teleportation of a $S$ gate~\cite{Bravyi05}.
More generally, the ability to update the Pauli frame in real time will be essential to implement quantum algorithms at the logical level. Since not all gates can be transversal in any given code~\cite{Eastin09}, conditional operations based on the current frame can be used to complete the universal gate set (e.g. $T$ gates in the surface code~\cite{Fowler12}.)

In conclusion, we have demonstrated the repeated measurement and real-time processing of stabilizers for a minimal bit-flip code. An intertwined readout and control system provides a real-time interface to the quantum processor,
converting a series of stabilizer results to the current Pauli frame without interrupting the execution of a potential algorithm. This approach is not limited to superconducting qubits, but is applicable to any quantum computing platform that faces coherence-limited operation.

Furthermore, this control architecture can be extended to larger-distance circuits such as the surface code. However, in the current implementation, the size of the lookup table limits the number of syndrome measurements that can be processed simultaneously~\cite{SOM}.
Future development will target the hardware implementation of the decoder~\cite{Devitt10}, possibly using neural networks to optimize  performance~\cite{Chamberland18a}.

\begin{acknowledgments}
We thank G.E.~Rowlands, D.~Ellard, and G.J.~Ribeill for their contributions to the software stack underlying this work, B. Hassick and A. Kreider for experimental assistance, M.~Takita, A.D.~C\'orcoles, B.~Abdo, J.M.~Chow for work on the qubit and Josephson amplifier devices. This research was funded by the Office of the Director of National Intelligence (ODNI), Intelligence Advanced Research Projects Activity (IARPA), through the Army Research Office Contract No. W911NF-14-1-0114.
\end{acknowledgments}

\renewcommand{\bibnumfmt}[1]{[S#1]}
\renewcommand{\citenumfont}[1]{S#1}
\renewcommand{\theequation}{S\arabic{equation}}
\renewcommand{\thefigure}{S\arabic{figure}}
\renewcommand{\thetable}{S\arabic{table}}
\setcounter{figure}{0}
\setcounter{equation}{0}
\setcounter{table}{0}

\section{Supplemental Material}

\subsection{Methods}

\label{sec:meth}
The real-time protocols presented in the manuscript rely on the interconnection between the receivers (two Innovative X6-1000M digitizers), the Processor (BBN Trigger Distribution Module -- TDM in Ref.~\cite{Ryan17S}), and the Pulse Sequencers (BBN Arbitrary Pulse Sequencers -- APS2).
The event sequence and the communication between those instruments for the experiment in Fig.~3 in the main text are illustrated below.

\begin{enumerate}[nolistsep]
  \item Processor initializes Decoder for $N$ cycles.
  \item Pulse Sequencers perform the set of the gates for the $n$th QEC cycle (all cycles are identical).
  \item Pulse Sequencers perform a measurement on the ancilla qubits $\At, \Ab$.
  \item The ancilla qubit Receiver (Digitizer X6-1000M (A) in Fig.~\ref{fig:S4}) converts the readout signals into assigned qubit states. The results $\{\at^n, \ab^n\}$ are fed back to the Processor, which stores them in the appropriate registers.
  \item Repeat from step 2 until the $N$ cycles are completed.
  \item Processor loads the stored syndromes $\{\at^n, \ab^n\}$ for $n=1...N$
  \item Decoding step: Processor calls a lookup table with the mapping between the aggregate syndrome measurements and the most likely bit flip errors~\cite{Tomita14S}. There are 8 possible decoder outputs, corresponding to up to one bit flip for each of the 3 data qubits (an even number of bit flips on the same qubit will cancel each other).
  \item Processor broadcasts the results of decoding to all the Pulse Sequencers.
  \item Pulse Sequencers issue the corrective $\hatX$ gates to the appropriate data qubits.
  \item Initialize Majority function.
  \item Measure the data qubits $\Da, \Db, \Dc$.
  \item The data qubit Receiver (Digitizer X6-1000M (B)) converts the readout signals into assigned qubit states. The results $\{\da, \db, \dc\}$ are fed back to the Processor, which stores them in the appropriate registers.
  \item Processor loads the stored $\{\da, \db, \dc\}$. These results are also saved on the measurement computer (Fig.~3b).
  \item Processor applies the majority function on the bits above.
  \item Processor broadcasts the result to all the Pulse Sequencers.
  \item A Pulse Sequencer issues a digital output when the majority output equals 1. This is acquired by one of the digitizers and saved on the measurement computer (Fig.~3c).
\end{enumerate}

\begin{figure*}
  \centering
  \includegraphics[width=1.7\columnwidth]{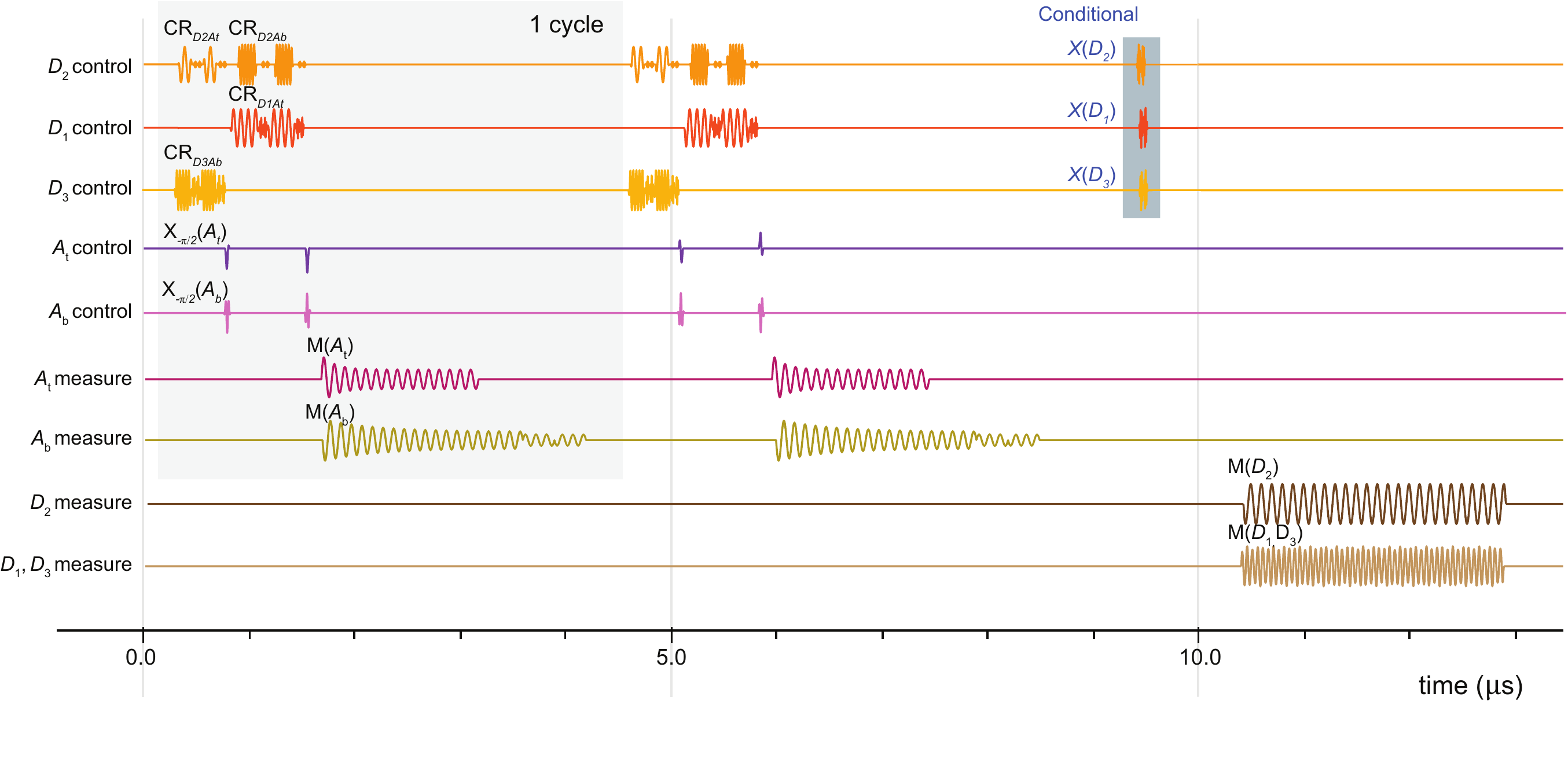}
  \caption{Detailed gate circuit for the active correction case of Fig.~3 and $N=2$. All pulses are generated by BBN APS2 pulse sequencers~\cite{Ryan17S} and upconverted with I/Q mixers to the desired qubit and cavity frequencies (see Table~\ref{table:1}). For simplicity, only one channel per I/Q pair is shown and state preparation is omitted.  Single-qubit gates are implemented with a diatomic Clifford library made of $X_{\pi/2}$ and partial $Z$ rotations~\cite{McKay17}.
  CNOT gates are implemented by concatenating echo cross-resonance (CR) pulses with the appropriate single-qubit gates~\cite{Chow14S}.
  For each qubit, a single channel is used to produce its own gates and all the CR tones where it plays the role of the control qubit. Thus, for example, the $D_2$ channel switches coherently between $f_{D2}$, $f_{At}$, and $f_{Ab}$ to implement the CNOT targeting $\At$ first, followed by one targeting $\Ab$ (see Figs.~1a, \ref{fig:S4}). This is achieved using the on-board fast frequency hopping developed in Ref.~\cite{Ryan17S}.
  Measurement pulses are also generated by one I/Q pair per qubit, with the exception of $\Da, \Dc$, whose pulses are combined and applied simultanously to both qubits.
  }
  \label{fig:S1}
\end{figure*}

\subsection{Classical hardware}
To facilitate this sequence, key modifications were made to the Pulse Sequencer and Processor gateware, relative to the system described in Ref.~\cite{Ryan17S}. In particular the `steering logic' was greatly expanded to enable the necessary Processor functions. The logic defines the relationship between measured inputs and control commands broadcast to the Pulse Sequencers.

First, acceleration engines were implemented on the Processor. The two engines currently available support the calculations necessary for the bit-flip code: Decoding and Majority. The Decoding lookup table stores a mapping of precomputed combinations relating syndrome measurements and data qubit errors. Currently the table supports up to eight cycles, with two ancilla measurements per cycle. The Majority function, here used on the final data qubit readout, can operate on up to 32 bits.

Second, to store measurement and calculation results, 4 kB of block RAM is allocated on the Processor FPGA.
This memory allows the Processor to accumulate measurement results to be used by the engines (Fig.~1c). Calculation results are also stored in this memory, at separate addresses from the measurement results.

Section IV of Ref.~\cite{Ryan17S} describes the original communication capability between Pulse Sequencers and Processor. Broadcasts from the Processor were limited to triggers and the results of one-time measurements.
Consequently, for this work, the one-way SATA communication link was modified to support additional data transfer functionality. The link now also supports address and data pair transfers (sent as an eight-byte message).
In this way, the Processor can transmit selected contents of its block RAM, such as engine results, to all Pulse Sequencers. In the case of active correction, a Pulse Sequencer utilizes this information to emit a pulse $590~\ns$ after the last measurement result is made available (Table~\ref{table:2}).

Control functions have been integrated into a publicly-available software stack~\cite{QGL} to initialize the steps in Sec.~\ref{sec:meth} and manage the address space in the Processor block RAM.

\begin{table}
\renewcommand{\arraystretch}{1.05}
\begin{ruledtabular}
\begin{tabular*}{\columnwidth} {l|c}

\multicolumn{1}{c}{Step}
 & \multicolumn{1}{r}{Latency (ns)} \\
\toprule
Receiver to Processor interface & 20  \\

Measurement storage in RAM & 50  \\

Engine initialization and calculation & 150  \\

Processor to PS module interface (8 bytes) & 210 \\

PS branching and waveform preparation & 130  \\
DAC output & 30 \\

Total & 590 \\

\end{tabular*}
\end{ruledtabular}

 \caption{\label{table:2} Additional latency after qubit state assignment. Latency includes the time to execute a Processor Engine, broadcast its results, and play corrective pulses from a Pulse Sequencer (PS).}

\end{table}

\begin{figure}
  \centering
  \includegraphics[width=0.9\columnwidth]{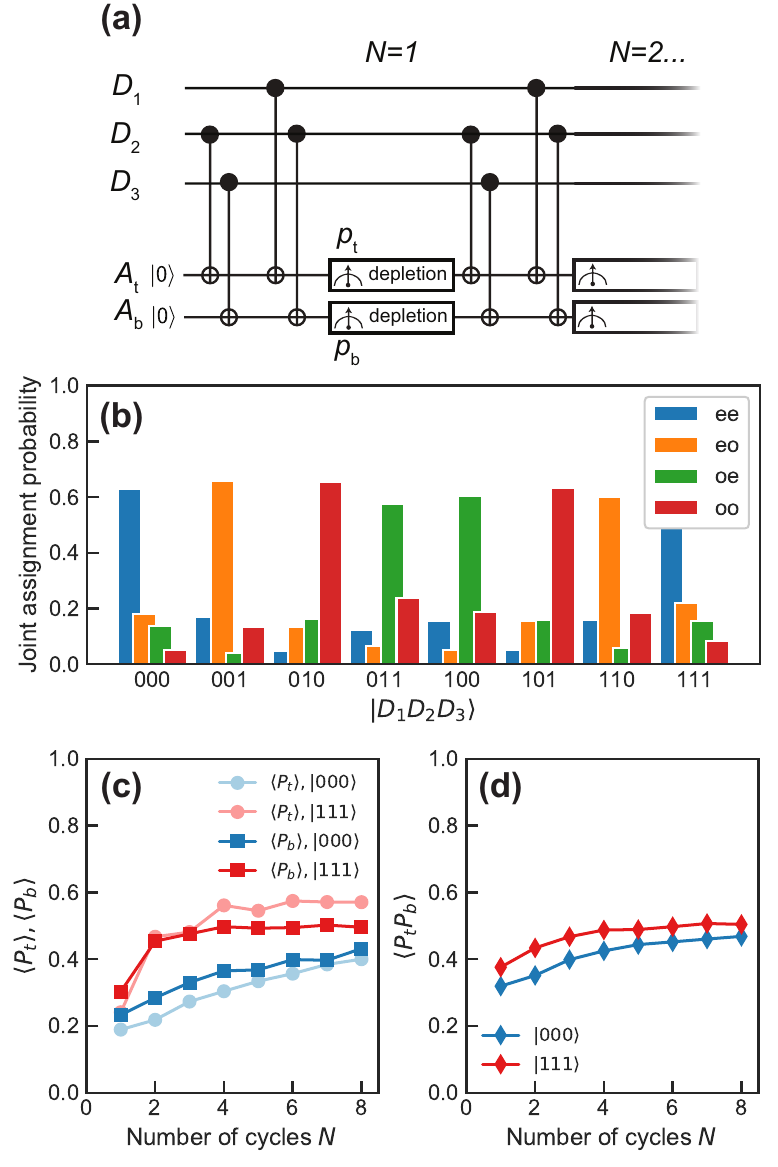}
  \caption{(a) Repeated stabilizer measurement. $p_{t(b)} = e/o$ is the single-shot parity assigned to the corresponding data qubit pair, with $e \leftrightarrow 0$ and $o \leftrightarrow 1$ indicating their relation to $a_{t(b)}$.
  (b) Probability to detect joint parity $ee$, $eo$, $oe$, and $oo$ for $N=1$ cycle of Fig.~1b and each of the 8 $\Da \Db \Dc$ computational states as input. (c) Average result of the individual parity measurements $a_t$ (circles) and $a_b$ (squares), as a function of number of cycles $N$, for initial states $|000\rangle$ (blue) and $|111\rangle$ (red). (d) Same as (c), but displaying the average of the joint parity measurement $a_t a_b$. Note that the ancillas are not reset between cycles.}
  \label{fig:S2}
\end{figure}

\begin{figure*}
  \centering
  \includegraphics[width=1.8\columnwidth]{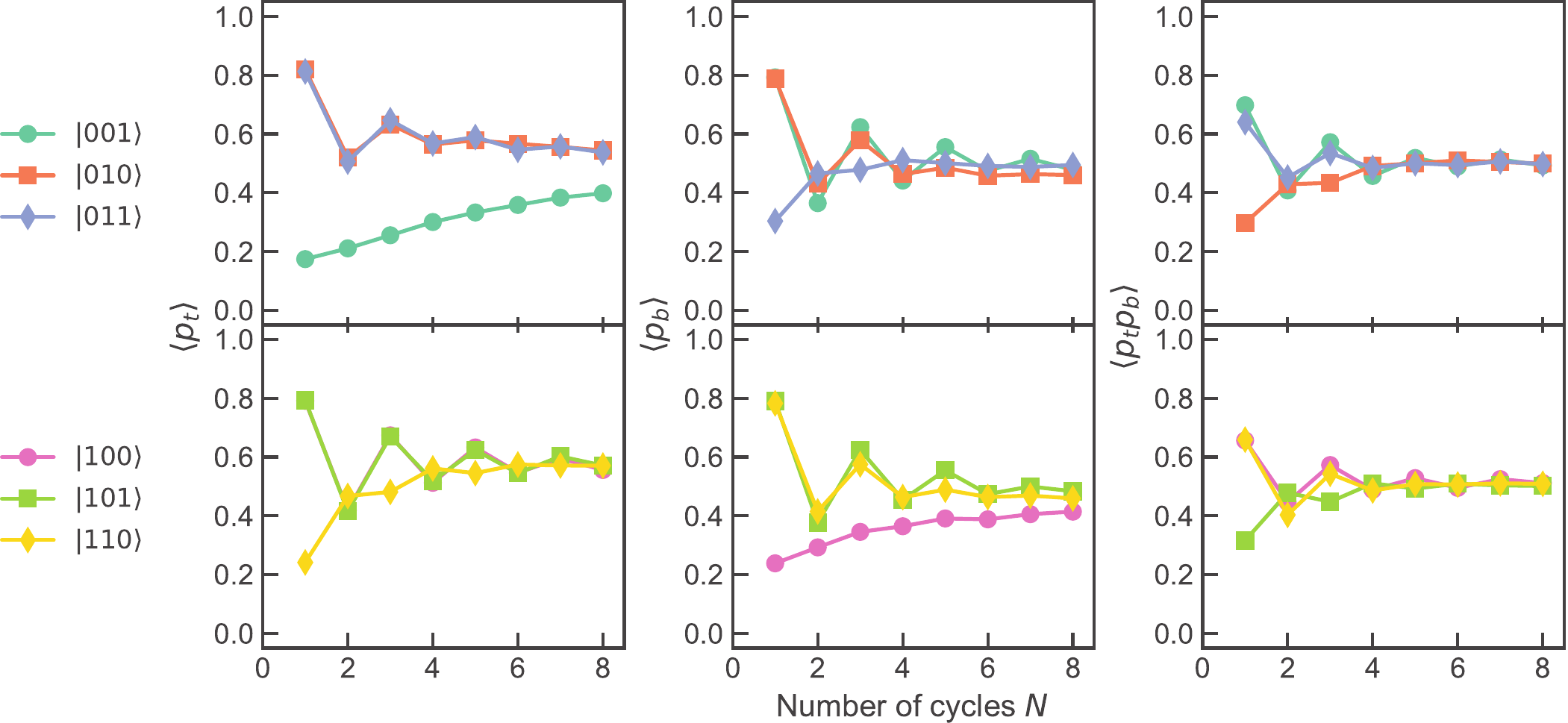}
  \caption{Repeated parity measurements for the computational states complementary to Fig.~\ref{fig:S2}.}
  \label{fig:S3}
\end{figure*}

\begin{table*}
\begin{ruledtabular}
\begin{tabular}{@{}ccccccccc@{}}

Qubit    & $f_{01}$ (GHz) & $T_2^\ast (\us)$  & $T_1 (\us)$ & $\eta_{kn}~\mathrm{(kHz)}$ & $\epsbar$ & $\epsilon_{1q}$ & $\epsilon_{2q}$ \\
\toprule
$\At$      & 5.313         & $26 \pm 8$                     & $48 \pm 9$            & \begin{tabular}{cc} 30 ($\eta_{t1}$);\\ 25 ($\eta_{t2}$) \end{tabular} & 0.075 &  0.003 & \begin{tabular}{cc} 0.07 ($\Da$);\\ 0.075 ($\Db$) \end{tabular} & \\
$\Ab$       & 5.362         & $39 \pm 6$                    & $49 \pm 2$            &  \begin{tabular}{cc} 8 ($\eta_{b2}$);\\ 35 ($\eta_{b3}$) \end{tabular} & 0.135 & 0.004  & \begin{tabular}{cc} 0.07 ($\Db$);\\ 0.035 ($\Dc$) \end{tabular}\\
$\Da$    & 5.412          & $12 \pm 4$                     & $29 \pm 4$        &  32 ($\eta_{12}$) & 0.25 & 0.0015 &  \\
$\Db$       & 5.220         &  $10 \pm 1$                  &  $7.7 \pm 0.6$ &   72 ($\eta_{23}$) & 0.13 & 0.01 &  \\
$\Dc$       & 5.408         & $38 \pm 11$                 &  $42 \pm 4$ &            & 0.22 & 0.001 &  \\

\end{tabular}
\end{ruledtabular}

\caption{\label{table:1} Device parameters. $\eta_{kn}$ indicate the static crosstalk terms in Eq.~\eqref{eqn:HCR}. $\epsbar$ is the single-qubit readout assignment error, averaged over the two states.
For each of $\Da,\Db,\Dc$, we also average this error over the state of the other two data qubits, to account for the crosstalk arising from their simultaneous measurement. The crosstalk between $\At$ and $\Ab$ is negligible. $\epsilon$ denote the average single- and two-qubit (target qubit between brackets) Clifford gate errors obtained by randomized benchmarking~\cite{Magesan12}.
This device was also used in Ref.~\onlinecite{Govia19} in a separate cooldown.}
\end{table*}

\begin{figure*}
  \centering
  \includegraphics[width=1.8\columnwidth]{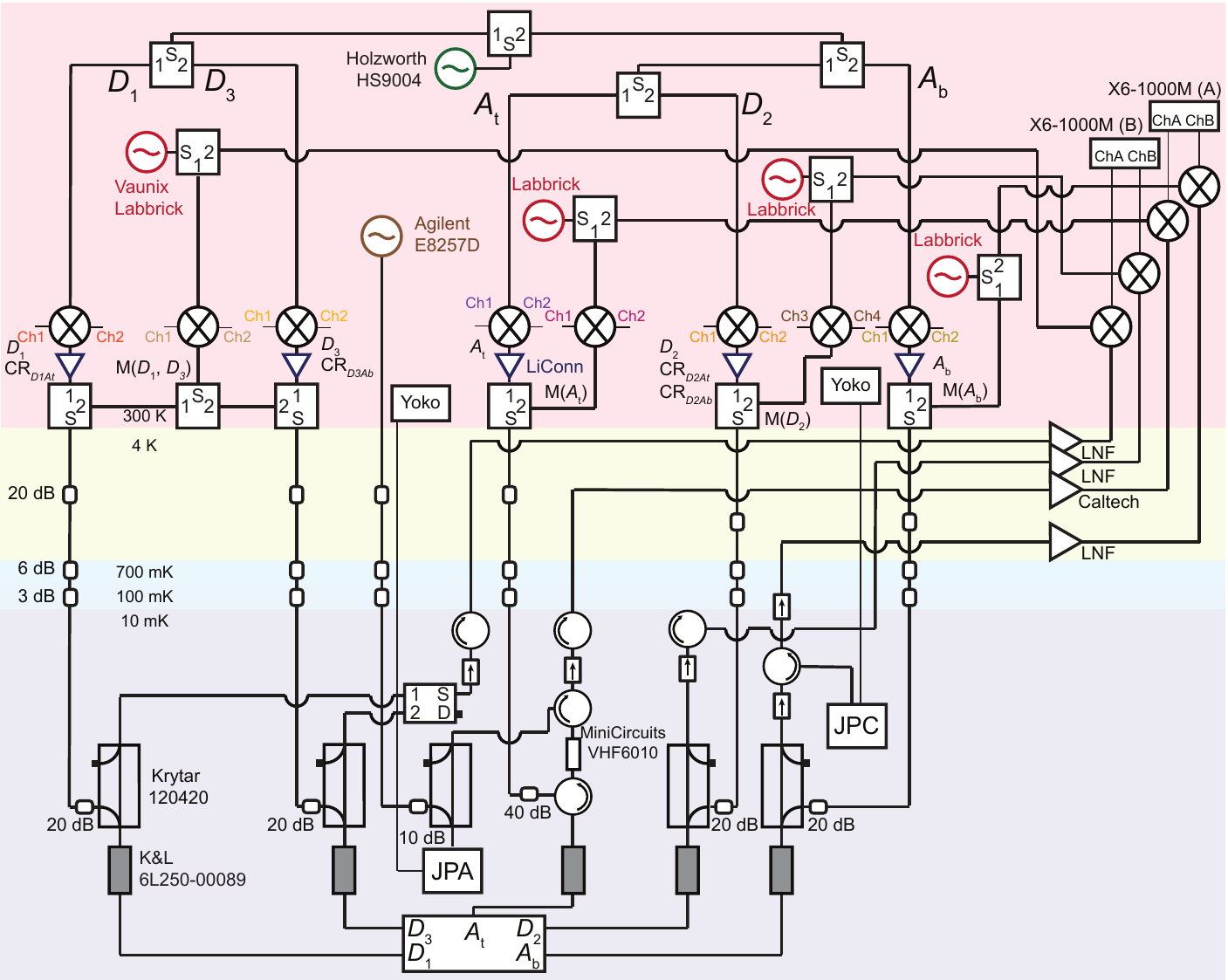}
  \caption{Detailed wiring from room temperature through the various stages of the Bluefors refrigerator housing the qubit device.
  The pulse waveforms produced by the Pulse Sequencers are input to the ports of IQ mixers and upconverted to qubit or resonator frequencies (see also Fig.~\ref{fig:S1}).}
  \label{fig:S4}
\end{figure*}

\section{Numerical Simulations}

In this section we describe the model and methods used to obtain the numerical simulation results shown in the main text. We chose not to do a full time-dependent master equation simulation of the error correction, as such an open system simulation is numerically intensive. Further, for the real-time error correction with ancilla reset, interspersing strong measurement and conditional operations within a time-dependent evolution is a nontrivial task. Instead, we use a simulation model that is approximate, but with well-controlled error that does not significantly reduce the accuracy of our results. We aim for qualitative agreement with the experimental results using a model with no fit parameters (i.e.~we do not search amongst models for a best fit to the data), with all free parameters in our model determined by independent characterization of the device.

Each round of the error correction can be thought of as an entangling operation (CNOT gates), followed by a measurement operation, followed by an optional correction and ancilla reset operation. This can be represented by the following composition of linear maps
\begin{align}
  \rho_{i+1} = \mathcal{R}_{\bf{m}}\circ\mathcal{M}\circ\mathcal{E}\left(\rho_i\right),
\end{align}
where $\rho_i$ is the state \emph{after} the $i$'th round of error correction, and $\mathcal{E}$, $\mathcal{M}$, and $\mathcal{R}_{\bf{m}}$ are the entangling, measurement, and correction/reset operations respectively, with the correction/reset operation being conditional on the vector of ancilla measurement outcomes $\bf{m}$. We now describe how we simulate each of these operations in detail.

\subsection{The Entangling Operation}

The ideal entangling operation is a sequence of CNOT gates between the data and ancilla qubits. We account for experimental imperfections in two ways, which roughly correspond to a separation of the coherent and incoherent error. We account for coherent error in the CNOT gates using one of two models. Our first model for coherent error in the CNOT gates is that there is some over-rotation in the $ZX$-interaction, and stray $ZZ$-coupling between connected pairs of qubits. The CNOT gate on data qubit $d$ and ancilla qubit $a$ is described by the unitary
\begin{align}
  \hat{U}_d^a = \hat{Z}_{90}^d\hat{X}_{90}^a \exp\left(-i\hat{H}_{d}^{a}t^a_d\right), \label{eqn:UCR}
\end{align}
where $\hat{Z}_{90}^d$ ($\hat{X}_{90}^a$) is a single qubit $90^\circ$ rotation on data qubit $d$ (ancilla qubit $a$) about the $Z$ ($X$) axis, which we assume is perfect. The cross-resonance interaction Hamiltonian is given by
\begin{align}
  \hat{H}_{d}^{a} = \frac{\pi/2 + \beta^d_a}{2t^a_d}\hat{Z}_d\otimes\hat{X}_a + \sum_{k,n\in \mathcal{C}}\eta_{kn}\hat{Z}_k\otimes\hat{Z}_n, \label{eqn:HCR}
\end{align}
where we use a compact notation for our operators that assumes an identity operator on a qubit unless otherwise written.

Here $t^a_d$ and $\eta_{kn}$ are the gate length, and experimentally characterized $ZZ$-coupling matrix. The set $\mathcal{C}$ describes all qubits that are connected by a resonator (such that a stray $ZZ$-coupling exists) but with connections to the data qubit $d$ removed. This is because we implement a single-echo CR gate \cite{Corcoles13,Takita16}, which cancels the leading order effect of stray $ZZ$-coupling with the data qubit. The angle $\beta^d_a$ characterizes the over-rotation error, which we estimate from 2-qubit randomized benchmarking (RB) of our CNOT gates. This is an overestimate of the rotation error, as both decoherence and $ZZ$-coupling also contribute to the RB decay. However, we take an overly cautious view of our device performance and attribute all of the RB decay to over-rotation error.

It is important to note that the Hamiltonian in Eq.~\eqref{eqn:HCR} is a simplification of the actual time-dependent Hamiltonian describing the fact that there is a pulse-shape associated with the CR drive. The dynamics generated by such a time-dependent Hamiltonian and Eq.~\eqref{eqn:HCR} are equivalent if there is no $ZZ$-coupling. When $ZZ$-coupling is present, the time-dependent version of Eq.~\eqref{eqn:HCR} no longer commutes with itself at different times, and the expression in Eq.~\eqref{eqn:UCR} is not the full transformation, but the leading order approximation via a Magnus expansion. While it is important to point out these (and other) discrepancies between theory and experiment, we do not believe that they have a significant enough effect to diminish the qualitative conclusions of our simulations.

Our second model for coherent error also uses the unitary of Eq.~\eqref{eqn:UCR} generated by Eq.~\eqref{eqn:HCR}, but with $\beta^d_a=0$ for all data-ancilla pairs (keeping the stray $ZZ$-coupling). Instead, we model the ancilla qubit error probabilistically, using the experimentally measured probabilities, $\gamma^a$, that either of the ancilla qubits have erroneously changed state after the entangling operation (i.e.~conditioned on no data qubit error). We draw a pseudorandom number $s$, and if $s < \gamma^a$ we say that an error has occurred and flip the state of the ancilla qubit. This probabilistic model for coherent error is in fact not coherent, but is rather the incoherent approximation to the coherent error sources that cause ancilla qubit flips. In our simulations, the ancilla flips are (potentially) implemented after the CNOT unitary operation.

We include incoherent error by introducing decoherence due to qubit relaxation ($T_1$) and dephasing ($T_2$). We follow our imperfect unitary with a decoherence operation $\mathcal{N}_{t^a_d}$, which we implement using the Kraus operator formalism~\cite{Nielsen10S}. This operation implements an amplitude and phase damping channel on \emph{all} qubits (data and ancilla) using the experimentally characterized values of $T_1$ and $T_2$ for a time duration $t^a_d$ that is the length of the CNOT gate on data qubit $d$ and ancilla qubit $a$.

The complete circuits for the entangling operation for both models of the coherent error are given by
\begin{align}
  &\mathcal{E}_1 = \mathcal{N}_{\tau_2}\circ\mathcal{U}_{2}^{2}\circ\mathcal{U}_{1}^{3}\circ\mathcal{N}_{\tau_1}\circ\mathcal{U}_{2}^{1}\circ\mathcal{U}_{1}^{2}, \\
  &\mathcal{E}_2 = \mathcal{A}_{F}\circ\mathcal{N}_{\tau_2}\circ\mathcal{U}_{2}^{2}\circ\mathcal{U}_{1}^{3}\circ\mathcal{N}_{\tau_1}\circ\mathcal{U}_{2}^{1}\circ\mathcal{U}_{1}^{2},~~~~\beta^a_d = 0,
\end{align}
where $\mathcal{U}_d^a(\rho) = \hat{U}_d^a\rho(\hat{U}_d^a)^\dagger$, and $\mathcal{A}_{F}$ is the probabilistic ancilla flip of coherent error model 2. This circuit encodes that fact that the CNOT gates are done simultaneously in pairs, and as such $\tau_1 = {\rm max}[t^2_1,t^1_2]$ and $\tau_2 = {\rm max}[t^3_1,t^2_2]$. In practice, we find minimal difference between the simulation results of the two circuits, and the results presented in the main text have used the second model of coherent error, given by $\mathcal{E}_2$.

This separation of decoherence from the unitary operation is an approximation, and a more physically accurate picture would arise from solving the time-independent master equation
\begin{align}
  \dot{\rho} = -i\left[\hat{H}_{d}^{a},\rho\right] +\sum_n\left( \frac{1}{T^{(n)}_1}\mathcal{D}\left[\hat{\sigma}_-\right] + \frac{1}{2T^{(n)}_\phi}\mathcal{D}[\hat{Z}]\right)\rho,
\end{align}
for a duration $t^j_k$ for the CNOT gate on qubits $d_j$ and $a_k$. Here $\mathcal{D}[\hat{x}]\rho = \hat{x}\rho\hat{x}^\dagger - \left\{\hat{x}^\dagger\hat{x},\rho\right\}/2$ is the usual dissipator, and $T_\phi = 2T_1T_2/(2T_1-T_2)$ is the pure dephasing rate. The summation indexed by $n$ is over all qubits (data and ancilla) which in general have different coherence times. Our decomposition into separate decoherence and unitary operations captures the dominant effects of the full simulation, and is accurate to the level we desire. It also considerably simplifies the composition of multiple CNOTs on different qubits into the full operation $\mathcal{E}$.

\subsection{The Measurement Operation}

To simulate strong quantum measurement we use a stochastic quantum simulation technique. After each entangling operation, we calculate the probability that each ancilla is in its excited state
\begin{align}
  p_e^{(a)} = {\rm Tr}\left[\ketbra{e}_a\mathcal{E}\left(\rho_i\right)\right].
\end{align}
We then generate two pseudorandom numbers, $r^{(a)}$. If $r^{(a)} < p_e^{(a)}$, we project ancilla $a$ to the excited state, and renormalize the density matrix. Otherwise, we project to the ground state and renormalize.

This approach introduces probabilistic trajectories to our simulation, and each set of states $\{\rho_i\}$ describes a round-by-round history of the system evolution for one particular trajectory. We perform this simulation many times to generate sufficient realizations of the stochastic evolution to calculate the average evolution
\begin{align}
  \bar{\rho}_i = \mathbb{E}\left[\rho_i^{k}\right] = \frac{1}{K}\sum_k \rho_i^{k}
\end{align}
where $\rho_i^{k}$ is the $k$'th realization of the simulation at the $i$'th round, and $K$ is the total number of realizations. In practice we have chosen $K=10^5$.

We introduce noise into the measurement operation in the following ways. We simulate decoherence during the measurement process by following or preceding the ancilla state projection by the phase and amplitude damping channel described in the previous section. As before, we implement this decoherence on all qubits, for the measurement time $t_m$, and our simulations show little noticeable difference between these orderings of the decoherence channel and the projection operation.

To simulate measurement infidelity we again use a stochastic technique. For each trajectory, at each round we store the measurement results for later use in our decoder during the correction operation. We draw a second pair of random numbers, and if either of these is less than the corresponding measurement infidelity, we flip the value of the stored result for that ancilla. We emphasize that we are \emph{not} flipping the state of the ancilla in $\rho_i^k$ but only of the stored classical bit representing the measurement outcome.

Finally, we also simulate measurement back-action on the measured ancilla qubit using a stochastic technique. As before, we draw a third pair of random numbers, and if either of these is less than the corresponding measurement back-action rate, we flip the state of the ancilla, by acting directly on $\rho_i^k$. For both the measurement infidelity and measurement back-action rates, we use values obtained from the experimental characterization of measurement we have performed for each ancilla qubit in our device.

This measurement back-action comes from effects not captured by the standard description of dispersive measurement in circuit QED, which is usually considered to be quantum non-demolition \cite{Blais04}. These effects can have a variety of sources \cite{Boissonneault08,Slichter12,Govia16,Govia15,Malekakhlagh18,Petrescu19}, and can be either incoherent or coherent. Our device characterization indicates that for our experiment these effects are likely incoherent, hence the model we have chosen.

\subsection{The Correction and Reset Operation}

The final operation in our simulations is (optional) correction of the data qubits, and (optional) reset of the ancilla qubits. For simulations with no error correction at all (which form the baseline control experiment), nothing occurs during this operation. For multi-round decoding using the decoder from  Ref.~\cite{Tomita14S}, the correction operation happens only after the final round. In this case, $\bf{m}$ is a $2\times N$ matrix containing the ancilla measurement results for $N$ rounds of error correction. The multi-round decoder takes as input $\bf{m}$, and outputs a correction operation, consisting of either identity or $\hatX$ gates applied to each data qubit, to apply to $\rho^k_N$ (as a reminder $k$ indexes the trajectory). We assume this correction operation can be done without error, as it need not be done in physical hardware, but only in software frame-tracking (see main text).

For repeated single-round error correction, the correction operation happens after every round, and is therefore applied to each $\rho_i^k$. Here, $\bf{m}$ is a $2\times1$ vector containing the ancilla measurement outcomes for only the $i$'th round. No decoder is necessary to determine the correction in this case, as the values of $\bf{m}$ determine which data qubits should be flipped. The operator $\hat{C}_1\otimes\hat{C}_2\otimes\hat{C}_3$ is applied to $\rho_i^k$, with each $\hat{C}_i\in\{\hat{\mathbb{I}},\hat{X}\}$. This is followed by a conditional flip (i.e.~reset) of the ancilla qubits, that depends on $\bf{m}$.

As the repeated single-round correction must be applied in hardware, we believe it is worthwhile to simulate error in the correction operation. We add additional decoherence before the correction gates, using the same phase and amplitude damping channel as before (on all qubits). This is to account for the lag time between measurement and correction, during which the hardware determines the appropriate correction. We also introduce errors into the the correction gate $\hat{X}$ in one of two ways. The first is an incoherent error model, where we assume that with some probability (which can be determined from the experimentally characterized $\hat{X}$ gate error) the $\hat{X}$ gate fails, and nothing happens to the qubit. As we have done for all other stochastic processes in our simulation, we implement this probabilistic failure using pseudorandom number sampling.

The second possible way to implement errors is by including the stray $ZZ$-interactions into the correction gate via the Hamiltonian
\begin{align}
  \hat{H}_{\rm cor} = \frac{\pi}{2t_{1q}}\hat{C}_1\hat{C}_2\hat{C}_3\hat{C}_4\hat{C}_5 + \sum_{k,n\in \mathcal{A}}\eta_{kn}\hat{Z}_k\hat{Z}_n, \label{eqn:ZZ}
\end{align}
where $\hat{C}_i\in\{\hat{\mathbb{I}},\hat{X}\}$ are the correction operators for the data and ancilla qubits, with $t_{1q}$ the single-qubit gate time. We have suppressed the tensor product symbols for brevity of notation. The set $\mathcal{A}$ consists of all connected qubits such that a $ZZ$-interaction exists between them. Our imperfect correction unitary is thus given by
\begin{align}
  \hat{U}_{\rm cor} = \exp\left(-i\hat{H}_{\rm cor}t_{1q}\right).
\end{align}
Following this unitary, we implement a final decoherence operation on all qubits for a length equal to the single-qubit gate time $t_{1q}$. We again find little quantitative difference between the simulation results with these two reset error models.

\end{document}